\documentclass[fleqn]{article}
\usepackage{espcrc2}
\usepackage{graphicx}
\usepackage[figuresright]{rotating}
\newcommand{\ag}{a_{3g}}
\newcommand{\ap}{a_{\gamma}}
\newcommand{\ac}{a_c}

\newcommand{\psip}{\psi^{\prime}}

\newcommand{\psipp}{\psi^{\prime \prime}}
\newcommand{\pspp}{\psi(3770)}
\newcommand{\jpsi}{J/\psi}

\newcommand{\DDbar}{D\overline{D}}
\newcommand{\DzDzbar}{D^0 \overline{D^0}}
\newcommand{\DD}{D^+ D^-}
\newcommand{\EE}{e^+e^-}

\newcommand{\KKSC}{K^{*+}K^-}
\newcommand{\KKSN}{K^{*0}\overline{K^0}}

\newcommand{\OP}{\omega\pi^0}

\newcommand{\jpsipp}{J/\psi \pi^+\pi^-}

\newcommand{\ra}{\rightarrow}

\newcommand{\rhopi}{\rho\pi}
\newcommand{\rhopin}{\rho^0 \pi^0}

\newcommand{\beq}{\begin{equation}}
\newcommand{\eeq}{\end{equation}}
\newcommand{\beqn}{\begin{eqnarray}}
\newcommand{\eeqn}{\end{eqnarray}}
\newcommand{\beqns}{\begin{eqnarray*}}
\newcommand{\eeqns}{\end{eqnarray*}}
\newcommand{\bfg}{\begin{figure}}
\newcommand{\efg}{\end{figure}}
\newcommand{\bitm}{\begin{itemize}}
\newcommand{\eitm}{\end{itemize}}
\newcommand{\bnum}{\begin{enumerate}}
\newcommand{\enum}{\end{enumerate}}
\newcommand{\btbl}{\begin{table}}
\newcommand{\etbl}{\end{table}}
\newcommand{\btbu}{\begin{tabular}}
\newcommand{\etbu}{\end{tabular}}

\title{Measuring $\psipp \rightarrow \rhopi$ in $\EE$ experiment}
\author{P.~Wang\address[IHEP]{Institute of High Energy Physics,
P.O.Box 918, Beijing 100039, China}
\thanks{Supported by 100 Talents Program of CAS (U-25)},
C.~Z.~Yuan \addressmark[IHEP],
X.~H.~Mo\addressmark[IHEP]$^,$\address[CCAST]{China Center of Advanced
Science and Technology, Beijing 100080, China} }

\date{\today}
\begin{document}

\begin{abstract}
In $S$- and $D$-wave mixing scheme, the branching ratio of $\psipp
\rightarrow \rhopi$ is estimated. Together with the continuum cross section
of $\rhopi$ estimated by form factor, the observed cross section of $\rhopi$ 
production at $\psipp$ in $\EE$ experiment is calculated taking into account the interference effect between the resonance and continuum amplitudes and the initial state radiative correction. The behavior of the cross section reveals that the disappearance of $\rhopi$ signal just indicates the existence of the corresponding branching ratio ${\cal B}_{\psipp \rightarrow \rhopi}$ at the order of $10^{-4}$. 
\vspace{1pc}
\end{abstract}
\maketitle

\section{Introduction}

The lowest charmonium resonance above the charmed particle production
threshold is $\pspp$ (shortened as $\psipp$) which 
provides a rich source of $\DzDzbar$ and $\DD$ pairs, as anticipated
theoretically~\cite{eichten}. However, non-$\DDbar$ ($non$-$charmed$
$final$ $state$) decay of $\psipp$ was studied theoretically
and searched experimentally almost two decades ago. The OZI violation
mechanism~\cite{berger} was utilized to understand the possibility of
non-$\DDbar$ decay of $\psipp$~\cite{lipkin}, and experimental investigations
involving noncharmed decay modes could be found in Ref.~\cite{zhuyn}. 

To explain the large $\Gamma_{ee}$ of $\psipp$, it is 
suggested~\cite{eichten2,kuang} that the mass eigenstates $\psi(3686)$ 
(shortened as $\psip$) and $\psipp$ are the mixtures of the $S$- and $D$-wave of 
charmonia, namely $\psi (2^3 S_1)$  state and 
$\psi (1^3 D_1)$ state. Recently it is proposed that such mixing gives possible
solution to the so-called ``$\rho\pi$ puzzle'' 
in $\psip$ and $\jpsi$ decays~\cite{rosnersd}. In this scheme
\beq
\begin{array}{l}
\langle\rhopi |\psip\rangle =\langle \rhopi | 2^3 S_1 \rangle \cos \theta
                  -\langle \rhopi | 1^3 D_1 \rangle \sin \theta~, \\
\langle\rhopi |\psipp\rangle=\langle \rhopi | 2^3 S_1 \rangle \sin \theta
                  +\langle \rhopi | 1^3 D_1 \rangle \cos \theta~,
\end{array}
\label{sdmix}
\eeq
where $\theta$ is the mixing angle between pure 
$\psi(2^3 S_1)$ and $\psi(1^3D_1)$ states and is fitted from the leptonic widths of $\psipp$ and $\psip$ to be either $(-27 \pm 2)^{\circ}$ or 
$(12 \pm 2)^{\circ}$~\cite{rosnersd}. 
The latter value of $\theta$ is consistent with the coupled channel 
estimates~\cite{eichten2,heikkila} and with the ratio of 
$\psip$ and $\psipp$ partial widths to $\jpsipp$~\cite{kuang,besppjp}. 
Hereafter, the discussions in this Letter are solely for the 
mixing angle $\theta =12^{\circ}$.

If the mixing and coupling of $\psip$ and $\psipp$
lead to complete cancellation of $\psip \rightarrow \rhopi$ 
decay ($\langle\rhopi |\psip\rangle= 0$), 
the missing $\rhopi$ decay mode of $\psip$ shows up instead as decay
mode of $\psipp$, enhanced by the factor $1/\sin^2 \theta$.
For $\theta= 12^{\circ}$, the $\psipp$ decay branching ratio~\cite{rosnersd} 
\beq
{\cal B}_{\psipp\rightarrow\rhopi}=(4.1\pm1.4)\times10^{-4}~~.
\label{brphi2}
\eeq 

With the resonance parameters of $\psipp$ from PDG2002~\cite{pdg}, the
total resonance cross section of $\psipp$ production at Born order is 
$$ \sigma_{\psipp}^{Born} =\frac{12 \pi }{ M^2_{\psipp} } \cdot {\cal B}_{ee}
                   = (11.6 \pm 1.8) \mbox{~nb}~. $$
Here $M_{\psipp}$ and ${\cal B}_{ee}$ are the mass
and $\EE$ branching ratio of $\psipp$.
With Eq.~(\ref{brphi2}), the Born order cross section of 
$\psipp \rightarrow \rhopi$ is 
$$\sigma_{\psipp \rightarrow \rhopi}^{Born} = (4.8 \pm 1.9) \mbox{~pb}~.$$

It is known that at $\sqrt{s}=M_{\psipp}$, the total continuum cross 
section, which is 13~nb, is larger than that of resonance.
Due to the OZI suppression, the total cross section of non-$\DDbar$ 
decay from the resonance is much smaller than that from the continuum. 
For an individual exclusive mode, the contribution from the continuum
process may be larger than or comparable with that from the resonance
decay. For the $\rhopi$ mode, the cross section of the resonance decay is 
more than three orders of magnitude smaller than the total continuum cross 
section, thus the contribution from the continuum and the
corresponding interference effect must be studied
carefully and taken into account in case of significant
modification of the experimentally observed cross section.

In the following sections, the Born order cross 
sections from the continuum and the resonance decays are given by virtue
of the form factor and the $S$- and $D$-wave mixing model, then
the experimental observable is calculated taking into account 
the radiative correction and experimental conditions.
Finally the dependence of the observed $\rhopi$ cross section on the phase between the OZI suppressed strong decay amplitude and the electromagnetic decay 
amplitude is discussed.

\section{Born order cross section of $\rhopi$}\label{bornsct}

In $\EE$ annihilation experiment at the charmonium resonance $\psipp$, 
there are three amplitudes responsible for $\rhopin$ final state\footnote{
Generally for certain final state $f$, three amplitudes describe the following 
three processes:
\[
\begin{array}{rl}
\ag:& \EE \rightarrow \psip,\psipp \rightarrow ggg \rightarrow f ~; \\
\ap:& \EE \rightarrow \psip,\psipp \rightarrow \gamma^* \rightarrow f ~;\\
\ac:& \EE \rightarrow  \gamma^* \rightarrow f ~.
\end{array}
\]
The first two processes are called resonance processes, and are denoted together as $\psip,\psipp \rightarrow f$ for short, while the third one is called continuum process and denoted as $ \EE \rightarrow f$ for short.
}: the continuum 
one-photon annihilation amplitude $\ac$, the electromagnetic decay amplitude
of the resonance $\ap$ and the OZI suppressed strong decay amplitude of 
the resonance $\ag$~\cite{wymz}:
$$A_{\rhopin}(s) = \ag (s)+\ap (s) +\ac (s)~~ .$$ 
As to electromagnetic interaction, the $a_c$ and $a_\gamma$ are related
to the $\rhopi$ form factor:
$$
\ac(s) =  {\cal F}_{\rhopin}(s)~~,
$$
and
$$
\ap (s) = B(s) \cdot {\cal F}_{\rhopin}(s)~~,
$$
with the notation
$$B(s) \equiv \frac{3\sqrt{s}\Gamma_{ee}/\alpha}
                   {s-M^2_{\psipp}+iM_{\psipp} \Gamma_t}~~,$$
where $\alpha$ is the QED fine structure constant, $\Gamma_t$ and $\Gamma_{ee}$ are the total width and $\EE$ partial width of $\psipp$. 
The strong decay amplitude can be parametrized in terms of its
relative phase ($\phi$) and relative strength (${\cal C}$) to the 
electromagnetic decay amplitude:
$$\ag (s) = {\cal C}e^{i\phi} \ap (s) ~~,$$
where ${\cal C}$ is taken to be real. 

Using $\cal C$, $\phi$ and ${\cal F}_{\rhopin}$, $A_{\rhopin}$ becomes~\cite{wymphase} 
\beq
A_{\rhopin} (s) = [({\cal C}e^{i\phi}+1)B(s)+1 ] \cdot {\cal F}_{\rhopin}(s)~~, 
\label{rhopieq}
\eeq
so the total $\rhopin$ cross section at Born order is 
\beqn
\lefteqn{\sigma^{Born}_{\rhopin}(s)= \frac{4\pi\alpha^2}{3s^{3/2}}  
|A_{\rhopin} (s) |^2 q_{\rhopin}^3~~,} \label{Born}
\eeqn
where $q_{\rhopin}$ is the three momentum of $\rho^0$ or $\pi^0$ 
in the final state. 

Since there is no experimental information on $\rhopi$ cross section for
the continuum process at resonance peak, the $\OP$ form factor is used for 
estimation. According to the SU(3) symmetry~\cite{haber},
\beq
{\cal F}_{\rhopin}(s) =\frac{1}{3} {\cal F}_{\OP}(s) ~.
\label{rpopn}
\eeq
${\cal F}_{\OP}$ is 
measured at $\sqrt{s}=M_{\psip}$ to be~\cite{wmyOP} 
$$
\left| \frac{{\cal F}_{\OP}(M_{\psip}^2)}{{\cal F}_{\OP}(0)} \right|
= (1.6 \pm 0.4)\times10^{-2}.
$$
This is in good agreement with the model dependent calculation in 
Ref.~\cite{Gerard}
$$
\left| \frac{{\cal F}_{\OP}(s)}{{\cal F}_{\OP}(0)} \right|
= \frac{(2\pi f_\pi)^2}{3s} ~~,
$$
where $f_\pi$ is the pion decay constant, or
\beq
|{\cal F}_{\OP}(s)|=\frac{0.531 ~\mbox{GeV}}{s}~,
\label{formop}
\eeq
by using the $\OP$ form factor at $Q^2=0$ from the crossed channel decay
$\omega\rightarrow\gamma\pi^0$.

With the form factor in Eq.~(\ref{formop}), the Born order continuum cross 
section of $\rhopi$ production at $\psipp$ resonance peak 
is~\footnote{Hereafter $\rhopin$ is used 
for one of the three different $\rhopi$ isospin states, 
and $\rhopi$ for the sum of them.}
$$\sigma^{Born}_{\EE \rightarrow \rhopi}=4.4~\hbox{pb}~.$$

For the resonance part,
\beq 
|({\cal C}e^{i\phi}+1) \cdot {\cal F}_{\rhopin}(M^2_{\psipp})|^2
\Gamma^0_{ee} M^2_{\psipp}
= | \langle \rhopin | \psipp \rangle |^2~,
\label{resamp}
\eeq 
where $\Gamma^0_{ee}$ is the $\EE$ partial width without vacuum polarization correction~\cite{tsaiys}. Starting from Eq.~(\ref{sdmix}), it can be acquired
$$\langle\rhopin |\psipp\rangle
              =\frac{\langle\rhopin | 2^3 S_1\rangle}{\sin\theta}
             - \langle\rhopin |\psip\rangle \tan \theta ~.$$
Since there could be an unknown phase between
$\langle\rhopin |2^3 S_1 \rangle$ and $\langle\rhopin | 1^3 D_1\rangle$,
or equivalently a phase (denoted as $\alpha$) between
$\langle\rhopin |2^3 S_1 \rangle$ and $\langle\rhopin | \psip \rangle$,
$| \langle\rhopin | \psipp \rangle |$ is constrained in a range.
With model-dependent estimation ${\cal B}_{\psip\rightarrow\rhopi}= 
(1.11\pm 0.87)\times 10^{-4}$~\cite{wymphase},
and $\langle\rhopin |2^3 S_1 \rangle$ in Ref.~\cite{rosnersd},
for $\theta=12^{\circ}$,
$$| \langle\rhopin | \psipp \rangle |^2=
( 1.8 \sim 5.2) \times 10^{-5} \mbox{~GeV~},$$
or equivalently
\beq
{\cal B}_{\psipp \rightarrow \rhopi}=
( 2.5 \sim 7.2 ) \times 10^{-4}~~,
\label{brrpa}
\eeq
which corresponds to the variation of $\alpha$ from $0^{\circ}$ to 
$180^{\circ}$. So the relation between ${\cal C}$ and $\phi$ could be derived 
from Eq.~(\ref{resamp}).

For a given ${\cal B}_{\psipp \rightarrow \rhopi}$,
according to Eqs.~(\ref{rhopieq}) and (\ref{Born}), 
the observed cross section depends on the interference pattern between the 
continuum one-photon amplitude and the $\psipp$ decay amplitude. 
In case of $\phi= \pm 90^\circ$, the maximum constructive or destructive 
interference between $\ag$ and $\ac$ happens at the resonance peak; 
while $\phi= 0^\circ$ or $180^\circ$ leads to constructive 
or destructive interference between $\ag$ and $\ap$. 

\section{Observed cross section of $\rhopi$}\label{obssct}

Due to the rapidly varying Breit-Wigner formula and the $\rhopi$ form factor as 
the center of mass energy changes, the observed cross section depends strongly 
on the initial state radiative correction which reduces the center of mass 
energy, and the invariant mass cut ($m_{cut}$) which removes the events produced by the initial state radiation. Taking these into account,
the observed cross section becomes
$$\sigma^{obs} (s)=\int \limits_{0}^{x_m} dx 
F(x,s) \frac{\sigma^{Born} (s(1-x))}{|1-\Pi (s(1-x))|^2}~,$$
where 
$$x_m=1-m^2_{cut}/s~.$$
$F(x,s)$ has been calculated to the accuracy of 
0.1\%~\cite{rad.1,rad.2,rad.3} and $\Pi(s)$ is the vacuum polarization 
factor~\cite{vacuum}.

It should be emphasized that the radiative correction modifies the 
Born order cross section in a profound way. Firstly, it shifts 
upward the maximum total cross section~\cite{zlshp} to the energy 
$M_{\psipp}+0.75$~MeV\footnote{In this Letter, it is assumed 
that the experiments take data at the energy which yields 
the maximum total cross section. The observed cross sections 
are calculated at this energy instead of the nominal resonance mass.}. 
Secondly, the radiative correction changes the Born order 
cross section significantly. For example, if $\phi=-90^\circ$ 
and ${\cal B}_{\psipp \rightarrow \rhopi} = 4.1 \times 10^{-4}$, 
$\sigma^{Born}_{\rhopi}=5.6 \times 10^{-3}$ pb, after radiative 
correction $\sigma^{obs}_{\rhopi}=0.31$ pb for $x_m=0.02$.

\begin{figure}[htb]
\centerline{\includegraphics[width=8.0cm,height=6.0cm]{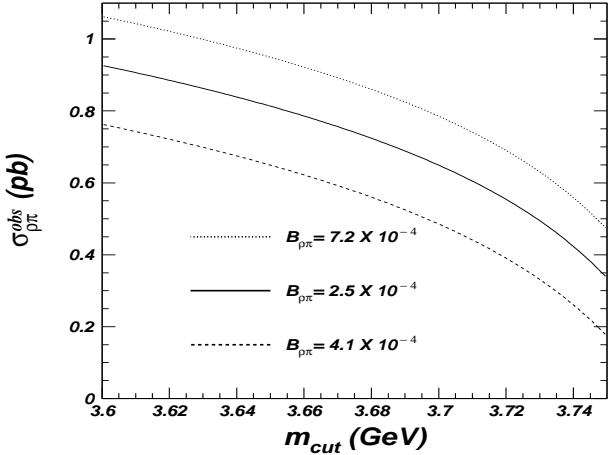}}
\caption{\label{ecutsg}The observed ${\rhopi}$ cross section as a function of 
$m_{cut}$ for three branching ratios ($\phi=-90^\circ$).}
\end{figure}

The dependence of the observed cross section on the invariant mass
cut is illustrated in Fig.~\ref{ecutsg}, for the branching ratio in
Eq.~(\ref{brrpa}) and $\phi=-90^\circ$.
It is obvious that a tighter invariant mass cut results in a smaller 
observed cross section. In the following analysis, $x_m=0.02$ is taken, 
which means a cut of $\rhopi$ invariant mass within 38~MeV from $M_{\psipp}$, 
or, near the midway between $\psip$ and $\psipp$ masses.

It is worth while to notice the variation of the observed $\rhopi$ cross section with the phase $\alpha$ between $\langle \rhopi | 2^3 S_1 \rangle$ and 
$\langle\rhopi | \psip \rangle$. When $\alpha$ varies from $0^{\circ}$ to 
$180^{\circ}$, $\sigma^{obs}_{\rhopi}$ in Fig.~\ref{ecutsg} moves from the line for ${\cal B}_{\psipp \rightarrow \rhopi} = 2.5 \times 10^{-4}$
to that for ${\cal B}_{\psipp \rightarrow \rhopi} = 4.1 \times 10^{-4}$,
and then increases to that for
${\cal B}_{\psipp \rightarrow \rhopi} = 7.2 \times 10^{-4}$,
at a specific invariant mass cut.

\begin{figure}[htb]
\centerline{\includegraphics[width=7.0cm,height=9.0cm]{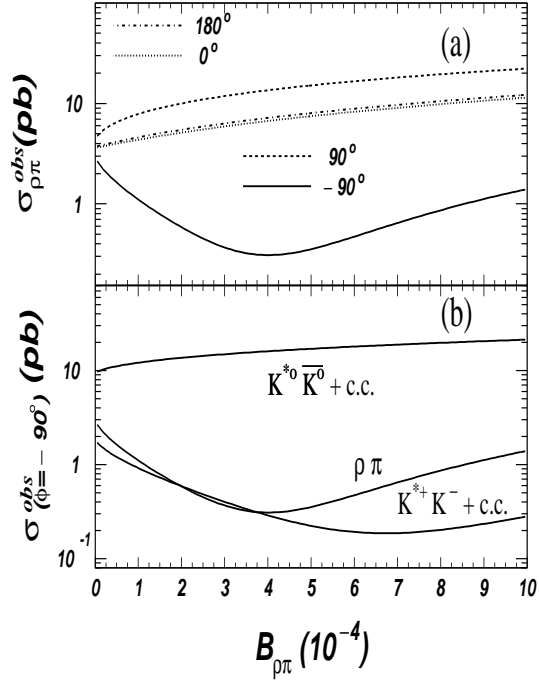}}
\caption{\label{brsect} (a) Observed $\rhopi$ cross section
as a function of ${\cal B}_{\psipp \rightarrow \rhopi}$ for different phases, 
and (b) observed cross sections of $\KKSN + c.c.$, $\KKSC + c.c.$, and 
$\rhopi$ as functions of ${\cal B}_{\psipp \rightarrow \rhopi}$ for 
$\phi=-90^{\circ}$.}
\end{figure}

The phase $\phi$ between $\ag$ and $\ap$ has significant effect on the observed cross section due to different interference patterns. 
Fig.~\ref{brsect}(a) shows the observed cross sections at $\psipp$ resonance 
peak as functions of ${\cal B}_{\psipp \rightarrow \rhopi}$ with $x_m=0.02$
and $\phi=-90^{\circ}$, $90^{\circ}$, $0^{\circ}$ and $180^{\circ}$, 
respectively. For the destructive interference between $\ag$ and $\ac$ 
($\phi=-90^{\circ}$), the cross section reaches its minimum for
${\cal B}_{\psipp \rightarrow \rhopi} \approx 4.1 \times 10^{-4}$, 
which corresponds to the resonance cross section of 3.3~pb,
but $\sigma^{obs}_{\rhopi}$ is only 0.31~pb, an order of magnitude
smaller. 

Above calculations of the observed cross section could be extended to other 
$1^- 0^-$ decay modes, such as $\KKSN+c.c.$ and $\KKSC+c.c.$, whose amplitudes 
are expressed as~\cite{wymphase}: 
\begin{eqnarray}
A_{\KKSN}=[({\cal C R} e^{i\phi}-2)B(s)-2 ] {\cal F}_{\rhopin} (s)~,
\label{kksneq}    \\
A_{\KKSC}=[({\cal C R} e^{i\phi}+1)B(s)+1 ] {\cal F}_{\rhopin}(s)~,  
\label{kksceq}    
\end{eqnarray}
where ${\cal R} \equiv (\ag + \epsilon)/\ag$, with $\epsilon$ describing 
the SU(3) breaking effect. It is assumed that $\epsilon$ has the same phase as 
$\ag$~\cite{suzuki}, so ${\cal R}$ is real. Using ${\cal C}$ determined from 
${\cal B}_{\psipp \rightarrow \rhopi}$ and ${\cal R}=0.775$ from fitting $\jpsi 
\rightarrow 1^- 0^-$ decay~\cite{wymphase}, the cross section of $\KKSN$ or 
$\KKSC$ is calculated by Eq.~(\ref{Born}) merely with the substitution 
of $A_{\KKSN}$ or $A_{\KKSC}$ for $A_{\rhopin}$. Their observed cross 
sections at $\psipp$ resonance peak as functions of
${\cal B}_{\psipp \rightarrow \rhopi}$ are shown in Fig.~\ref{brsect}(b) for 
$\phi=-90^{\circ}$ and $x_m=0.02$. It could be seen that the cross section of 
$\KKSN+c.c.$ is much larger than those of $\rhopi$ and $\KKSC+c.c.$ in a wide
range of the $\rhopi$ branching ratio.

\begin{figure}[htb]
\centerline{\includegraphics[width=8.0cm,height=8.0cm]{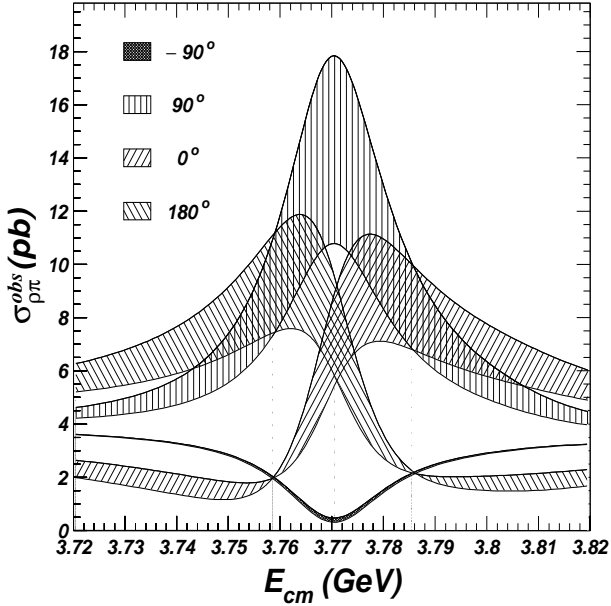}}
\caption{\label{ecmxct}Observed $\rhopi$ cross section varies with 
the center of mass energy for different phases: $\phi=-90^{\circ}$, 
$+90^{\circ}$, $0^{\circ}$, and $180^{\circ}$, respectively.}
\end{figure}

Since the data at $\psipp$ resonance peak alone can not fix all parameters 
(${\cal C}$, $\phi$, and ${\cal F}_{\rhopi}$) in 
${\cal B}_{\psipp \rightarrow \rhopi}$ determination, 
the correct way of measuring the branching ratio is through energy 
scan of the resonance. Fig.~\ref{ecmxct} shows the observed $\rhopi$ cross section
in the vicinity of the $\psipp$ resonance, with $x_m=0.02$ and the
branching ratio in Eq.~(\ref{brrpa}) for four values of $\phi$:
$-90^\circ$, $+90^\circ$, $0^\circ$, and $180^\circ$. The hatched areas
are due to the variation of $\alpha$. 
For $\phi=0^\circ$ or $180^\circ$, the maximum observed cross section is
above or below the resonance mass. Only for $\phi=+90^\circ$ the maximum 
observed cross section is near the resonance peak. Here the most 
interesting phenomenon is, with $\phi=-90^\circ$, the observed cross 
section reaches its minimum near the resonance peak! This phenomenon suggests 
that at the resonance peak the undetectable experiment cross section of 
$\rhopi$ just indicates the existence of the corresponding branching ratio at 
the order of $10^{-4}$.

\section{Discussion}\label{diss}

As shown in Fig.~\ref{ecmxct}, the line shape of the $\rhopi$ cross section is 
sensitive to the phase $\phi$. If the fine scan is infeasible, at least at 
three energy points the data must be taken in order to fix the three 
parameters in Eq.~(\ref{rhopieq}): $\cal C$, $\phi$ and ${\cal F}_{\rhopin}$.
According to Eq.~(\ref{sdmix}), with $\rhopi$ branching ratios at $\psip$ and 
$\psipp$, and the magnitude of $\langle \rhopi | 2^3 S_1\rangle$, 
which is derived in Ref.~\cite{rosnersd}, the relative phase 
$\alpha$ could also be determined. If only the data at $\psipp$ peak is 
available, a model-dependent way to determine the $\rhopi$ branching 
ratio is to look for more $1^-0^-$ modes, such as $\KKSN+c.c.$, $\KKSC+c.c.$ 
and $\OP$. Notice in Eqs.~(\ref{rhopieq}), (\ref{rpopn}), (\ref{kksneq}) and 
(\ref{kksceq}), the three modes are parametrized by four parameters: 
$\cal C$, $\cal R$, $\phi$ and ${\cal F}_{\OP}$. With the measurement of 
${\cal F}_{\OP}$ through $\OP$ mode, the other three parameters could be 
solved from Eqs.~(\ref{rhopieq}), (\ref{kksneq}) and (\ref{kksceq}).

Eqs.~(\ref{rhopieq}) and (\ref{kksneq}) indicate opposite interference 
patterns between $\ag$ and $\ac$ for $\rhopin$ and $\KKSN$. That is, if the interference between $\ag$ and $\ac$ for $\rhopi$ is {\bf destructive}, then 
such interference is just {\bf constructive} for $\KKSN+c.c.$ and $vice~ versa$.  In the resonance scan, if $\sigma^{obs}_{\rhopi}$ reaches its valley near
$\psipp$ resonant mass, $\sigma^{obs}_{\KKSN+c.c.}$ reaches its peak\footnote{In Fig.~\ref{ecmxct}, if the scan behavior of $\sigma^{obs}_{\rhopi}$ is similar 
to the curve corresponding to $\phi=-90^{\circ}$, $\sigma^{obs}_{\KKSN+c.c.}$ 
would be similar to the $\rhopi$ curve corresponding to $\phi=+90^{\circ}$.}.
This means if the observed $\rhopi$ cross section at $\psipp$ is 
smaller than that at continuum, the observed $\KKSN+c.c.$ cross section 
at $\psipp$ will be larger. So the measurements of $\KKSN+c.c.$ and 
$\rhopi$ provide a crucial test of the interference pattern between 
$\ag$ and $\ac$.

There are theoretical arguments in favor of the orthogonality between $\ag$ and $\ap$~\cite{gerarda} of the charmonium decays. The phenomenological 
analyses for many two-body decay modes: $1^-0^-$, $0^-0^-$, $1^-1^-$, $1^+0^-$,
and Nucleon anti-Nucleon on $\jpsi$ data support this 
assumption~\cite{suzuki,kopke}. The recent analysis of $\psip \ra 1^-0^-$ 
decays which took into account the contribution from the 
continuum, found that the phase $\phi=-90^\circ$ could fit current
available data within experimental uncertainties and $\phi=+90^\circ$ could be 
ruled out~\cite{wymphase}. Similar analysis of $\psip \ra 0^-0^-$ decays also 
favors the orthogonal phase~\cite{ywmphase}. For $\psipp$ , 
it is of great interest here to note 
the search of $\rhopi$ mode by MARK-III~\cite{zhuyn} at the $\psipp$ peak. 
The result corresponds to the upper limit of the $\rhopi$ 
production cross section of 6.3~pb at 90\% C. L., which favors $\phi=-90^\circ$ than other possibilities as seen from Fig.~\ref{ecmxct}. These experimental 
information suggests the phase $\phi=-90^\circ$ between $\ag$ and $\ap$ be
universal for all quarkonia decays. 

At last, a few words about the effect of the beam energy spread $\Delta$ for
cross section measurement~\cite{wmyOP}. $\psipp$ is a relatively wide resonance,
for a collider with small energy spread, such as $\Delta=1.4$~MeV at 
BES/BEPC~\cite{besdelta}, this effect is negligible. With increasing $\Delta$, 
the correction becomes larger. For example, on a collider with $\Delta=5$~MeV, 
for $\phi=-90^\circ$ and $x_m=0.02$, the observed cross section of $\rhopi$ for ${\cal B}_{\psipp \rightarrow \rhopi}=4.1 \times 10^{-4}$ is more than doubled to $0.68$~pb comparing with the value without energy spread effect.
\section{Summary}

By virtue of $S$- and $D$-wave mixing model, 
${\cal B}_{\psipp \rightarrow \rhopi}$ is estimated, 
together with the estimated $\sigma_{\EE \rightarrow \rhopi}$ by form
factor, the observed $\rhopi$ cross section at $\psipp$ in $\EE$ experiment, 
which takes into account the initial state radiative correction, 
has been calculated. The study shows that the disappearance of $\rhopi$ cross 
section at $\psipp$ peak just indicates the branching ratio of 
$\psipp \rightarrow \rhopi$ at the order of $10^{-4}$.

Besides the information of $\psipp$, if the phase analyses of $\jpsi$ and 
$\psip$ are also taken into consideration, it is natural to conclude that
the  phase $\phi=-90^{\circ}$ between $\ag$ and $\ap$ is universal for all quarkonia decays.

In the forthcoming high luminosity experiments of $\psipp$ at
CLEO-c ~\cite{cleoc} and BES-III~\cite{bes3}, the property of 
$\rhopi$ decay and the feature of the
phase are expected to be tested quantitatively.

\end{document}